\documentclass[aps,prd,twocolumn,superscriptaddress,nofootinbib,floatfix,noshowpacs]{revtex4-2}
\pdfoutput=1
\newif\ifcomment
\usepackage{amsmath,amssymb}
\usepackage{color,hyperref,url}
\usepackage{listings}
\usepackage{slashed}
\usepackage[pdftex]{graphicx}
\usepackage{epstopdf}
\usepackage{epsfig}
\usepackage{grffile}
\usepackage{relsize}
\usepackage{xcolor}
\graphicspath{{./img/}}
\usepackage{soul}
\newcommand{\beq}{\begin{equation}}
\newcommand{\eeq}{\end{equation}}
\newcommand{\ba}{\begin{array}}
\newcommand{\ea}{\end{array}}
\newcommand{\bea}{\begin{align}}
\newcommand{\eea}{\end{align}}
\newcommand{\bi}{\begin{itemize}}
\newcommand{\ei}{\end{itemize}}
\newcommand{\ben}{\begin{enumerate}}
\newcommand{\een}{\end{enumerate}}
\newcommand{\bc}{\begin{center}}
\newcommand{\ec}{\end{center}}
\newcommand{\bl}{\begin{flushleft}}
\newcommand{\el}{\end{flushleft}}
\newcommand{\br}{\begin{flushright}}
\newcommand{\er}{\end{flushright}}

\begin{document}

\preprint{APS/123-QED}

\title{Reconstructing parton distribution function based on maximum entropy method}% Force line breaks with \\

\author{Sihan Zhang}
\email{zhangsihan@mail.nankai.edu.cn}
\affiliation{%
College of Chemistry, Nankai University, Tianjin 300071, China 
}%
%\altaffiliation{College of Chemistry, Nankai University.}%Lines break automatically or can be forced with \\
\author{Xiaobin Wang}
\email{wangxiaobin@mail.nankai.edu.cn}
\author{Tao Lin} 
\email{lintaophy@mail.nankai.edu.cn}
\author{Lei Chang}
\email{leichang@nankai.edu.cn}

\affiliation{%
 School of Physics, Nankai University, Tianjin 300071, China 
}%

\date{\today}% It is always \today, today,
\begin{abstract}
A new method based on the maximum entropy principle for reconstructing the parton distribution function (PDF) from moments is proposed. Unlike traditional methods, the new method no longer needs to introduce any artificial assumptions. For the case of moments with errors, we introduce Gaussian functions to soften the constraints of moments. Through a series of tests, the effectiveness and reconstruction efficiency of this new method are evaluated comprehensively. And these tests indicate that this method is reasonable and can achieve high-quality reconstruction with at least the first six moments as input. Finally, we select a set of lattice QCD results regarding moments as input and provide reasonable reconstruction results for the pion.
\end{abstract}

\maketitle
%\tableofcontents
\section{\label{introduction}introduction}
At high energy, the scattering process with a hadron actually happens on its internal constituents, namely the quarks and the gluons, which are commonly called the partons. So if we want to get the scattering cross-sections of these scattering processes, the information from parton is necessary. And we usually describe these partons by using the parton distribution function (PDF), which is the probability that the parton carries a certain momentum fraction of the hadron momentum. Therefore, the determination of the PDFs of hadrons has always been an important project in hadron physics.

It is not easy to predict the PDF in theory since it will involve non-perturbative QCD. Traditionally, we can only calculate the first few moments of a PDF and then use some methods to reconstruct the PDF~\cite{PhysRevD.56.2743,Broemmel:2008pD,Javadi-Motaghi:201407,PhysRevD.99.014508,baron2007moments,PhysRevD.100.114512,PhysRevD.103.014508}. In recent years, there have been some new methods~\cite{Braun:2007wv,PhysRevLett.110.262002,Ji:2014gla,Radyushkin:2016hsy,PhysRevLett.118.242001} that can directly obtain PDF, but these methods still have many problems, such as excessive error and limited computable regions. So how to reconstruct PDF with finite moments is a problem that needs to be frequently addressed. Therefore, people have developed many different methods. But these methods always have to assume the functional form of PDF in advance and then use moments to determine the parameters, ultimately completing the reconstruction. The presupposition of the PDF functional form inevitably includes some artificial choices, which makes the reconstruction results not convincing enough.

In this work, we propose a new reconstruction method for the symmetric PDF of the pion to avoid the impact of artificial choice. This method will obtain the PDF by maximizing entropy under moment constraints, and we will not introduce any artificial presets about the PDF. This paper is organized as follows: Section~\ref{level1-1} introduces the detailed content of this new reconstruction method, including the cases of precise moments and moments with errors. Section~\ref{3} presents a series of calculation results, including validity test, reconstruction efficiency assessment base on artificial inputs, and the results corresponding to real inputs. A summary is presented in Section~\ref{4}.
\section{\label{level1-1} Model and Method}
\subsection{\label{sec:level2} Algorithm Design}

The maximum entropy method is utilized to determine the distribution function $f(x)$, which is a mature idea that has been practiced multiple times~\cite{Wang:2014lua,Han:2018wsw,Han:2020vjp}. The Shannon entropy of the distribution function $f(x)$ is defined as:
\begin{equation}
    S=-\int_{0}^1 f(x)\log f(x)dx .
\end{equation}

If the system has no constraints, it will produce a constant distribution function, which is the principle of equal a priori probabilities. However, the distribution of partons requires several constraints to be considered. These constraints for the pion at the hadron scale can be mathematically expressed as follows:
\begin{equation}
    \int _{0}^{1}f(x)dx=1,
\end{equation}
\begin{equation}
    f(x)=f(1-x),
\end{equation}
\begin{equation}
    \int _{0}^{1}x^i f(x)dx=\mu_{i}.
\end{equation}

We define a Lagrange function $L$ to describe the system so that entropy and all the constraints can be combined as follows:
\begin{equation}
    L = S+\sum_{i=0}^{m}\lambda_{i}\Delta_i,
\end{equation}
where $\lambda_i$ are unknown coefficients, and
\begin{equation}
    \Delta_i = \int_0^1x^if(x)dx - \mu_i^{\rm prior}
\end{equation}
where $\mu_i^{\rm prior}$ are the prior informations or constraints given by the first principle theory.

Given the challenges of obtaining an analytical solution for the distribution function, it is often beneficial to employ an approximate distribution function with undetermined coefficients. The more coefficients you choose, the more flexible the distribution function is. Additionally, taking into account the system's symmetry, we have selected a suitable basis set as follows:
\begin{equation}
    f(x)=\sum_{k=1}^{n}a_{k}\sin{(2k-1)}\pi x .
\end{equation}

If the system function has reached its maximum point, the derivation of the Lagrange function should be equal to zero:
\begin{equation}
    \frac{\partial L}{\partial\lambda_i}=0,\ \frac{\partial L}{\partial a_{j}}=0.
\end{equation}

Therefore, this reconstruction has been converted to the question of finding the solution of derivation equations. However, it is hard to find analytical solutions to these complex equations. To resolve this, we utilize the Self-Consistent Field (SCF) method, originally introduced by Hartree for solving multiple-electron systems~\cite{hartree_1928}. This method is valuable for seeking the steady status of a complex system with constraints and one system function. Starting with initial values, the SCF process iteratively finds nearby steady solutions. The SCF process can be described by using the recurrence equation:
\begin{equation}
    \begin{bmatrix} 
        \lambda_0\\
        \vdots\\
        \lambda_m\\
        a_1\\
        \vdots\\
        a_n
    \end{bmatrix}_{k+1}=
    \begin{bmatrix} 
        \lambda_0\\
        \vdots\\
        \lambda_m\\
        a_1\\
        \vdots\\
        a_n
    \end{bmatrix}_{k}+\xi\pmb{H}^{-1}
    \begin{bmatrix}
        -\Delta_0\\ \vdots \\ -\Delta_m\\
        -\frac{\partial L}{\partial a_1} \\ \vdots\\ 
        -\frac{\partial L}{\partial a_n}
    \end{bmatrix},
\end{equation}

In this equation, $\pmb{H}$ represents the bordered Hessian matrix of the Shannon entropy $S$. Specifically, it can be represented as:
\begin{equation}
    \pmb{H} =
    \begin{bmatrix} 
        0&\cdots&0& \frac{\partial \Delta_0}{\partial a_1}&\cdots  & \frac{\partial \Delta_m}{\partial a_1}
        \\
        \vdots&\ddots &\vdots &\vdots & \ddots &\vdots 
        \\
        0&\cdots&0 &\frac{\partial \Delta_0}{\partial a_n}&\cdots & \frac{\partial \Delta_m}{\partial a_n}
        \\
        \frac{\partial \Delta_0}{\partial a_1}&\cdots &\frac{\partial \Delta_0}{\partial a_n}&\frac{\partial^2 L}{\partial a_1\partial a_1} &\cdots &\frac{\partial^2 L}{\partial a_n\partial a_1}
        \\
        \vdots&\ddots&\vdots&\vdots&\ddots&\vdots  
        \\
        \frac{\partial \Delta_m}{\partial a_1}&\cdots &\frac{\partial \Delta_m}{\partial a_n}&\frac{\partial^2 L}{\partial a_1\partial a_n} &\cdots &\frac{\partial^2 L}{\partial a_n\partial a_n}
    \end{bmatrix},
\end{equation}
where
\begin{equation}
    \frac{\partial \Delta_m}{\partial a_k} = \int_0^1 x^m \sin{(2k-1) \pi x} dx,
\end{equation}
\begin{equation}
    \frac{\partial^2 L}{\partial a_j\partial a_k} = -\int_0^1 \frac{(\sin{(2j-1) \pi x})( \sin{(2k-1) \pi x})}{f(x)} dx.
\end{equation}

In this equation, $\xi$ represents the step size. A larger $\xi$ leads to a faster convergence process but may compromise program stability when the initial values are too bad. For all calculations in this paper, the default value of $\xi=1$ was used. Moreover, we set the iterative tolerance to $10^{-5}$, which provides sufficient accuracy for our calculations.

After obtaining the coefficients, a reasonableness test is necessary to ensure that the result represents the local maximum point of entropy. For this purpose, the bordered Hessian matrix must satisfy a sufficient condition: the leading principal minors starting from $2m+1$ must alternate in sign, with the smallest one having the sign of $(-1)^{m+1}$.

In conclusion, this SCF method contains five steps:
\begin{itemize}
    \item Guess the initial values of the coefficient array;
    \item Obtain the Hessian matrix (Integration is calculated by the grid point method);
    \item Calculate the inverse of the Hessian matrix and update the coefficient array;
    \item Determine whether the variation of the Lagrange function is less than the preset tolerance $10^{-5}$. If not, go back to the second step;
    \item Reasonableness test. If the test fails, repeat the process with the new initial value.
\end{itemize}

\subsection{\label{sec:citeref} Reconstruction from Constraints with Errors}
The calculated results of the moment of PDF by QCD are always accompanied by errors, represented as $\mu_i \pm \sigma_i$. Consequently, using the simple Lagrange functions to address these constraints is insufficient, as they impose excessively strict conditions. Therefore, it is necessary to relax the moment constraints. One approach to achieving this is by replacing the original constraint terms with a relaxation function. In our paper, we employ the Gaussian-shaped function as this relaxation function:
\begin{equation}
    \lambda_i\Delta_i\to E_i(f)=\frac{1}{\sqrt{2\pi}\sigma_i}\exp{\left(-\frac{\Delta_i^2}{2\sigma_i^2}\right)}\ (i=1,2,\dots,m).
\end{equation}

Notably, the above replacement starts from $i=1$ since $\Delta_0$, which serves as the normalization factor, is known to be completely accurate. The peak of the Gaussian-shaped function corresponds to the center of the error bar, with its value gradually increasing as $\Delta_i$ approaches zero. Additionally, as $\sigma_i$ increases to represent larger errors, the curve becomes flatter. Consequently, the constraint of moments is weakened in proportion to its uncertainty. Considering these properties, our replacement is a reasonable approach.

By introducing this Gaussian shape function, the new Lagrange function is corrected as follows:
\begin{equation}
    L' = S + \lambda_0 \Delta_0 + \beta \sum_{i=1}^m E_i(f),
\end{equation} 
where the coefficient $\beta$ represents the strength of the constraints. By adjusting the value of $\beta$, the model can effectively control the trade-off between maximizing the entropy and satisfying the moment constraints. A higher value of $\beta$ emphasizes the importance of meeting the constraints, resulting in a distribution that closely aligns with the specified moments. Conversely, a lower value of $\beta$ places more emphasis on maximizing the entropy, allowing for a distribution that may deviate slightly from the constraints. Similarly, the SCF method changed correspondingly by replacing $L$ with $L'$.

\section{Results}\label{3}
\subsection{Parameter Determination and Validity Assessment}

Within our model, two parameters, the term number of the basis set and the size of the grid point, necessitate determination through comparison with the analytical solution. In scenarios where the distribution is constrained solely by the second-order moment, the PDF can be ascertained by the variational approach. This approach yields a Gaussian function as the analytical solution. Specifically, when the second moment $\mu_2=0.3$, the Gaussian function is represented as follows, with an associated entropy of $-0.114$:
\begin{equation}
    f(x) = 1.63\exp{\left(-7.5(x-0.5)^2\right)}.
\end{equation}

\begin{figure}[!htbp]
    \centering
    \includegraphics[width=1\linewidth]{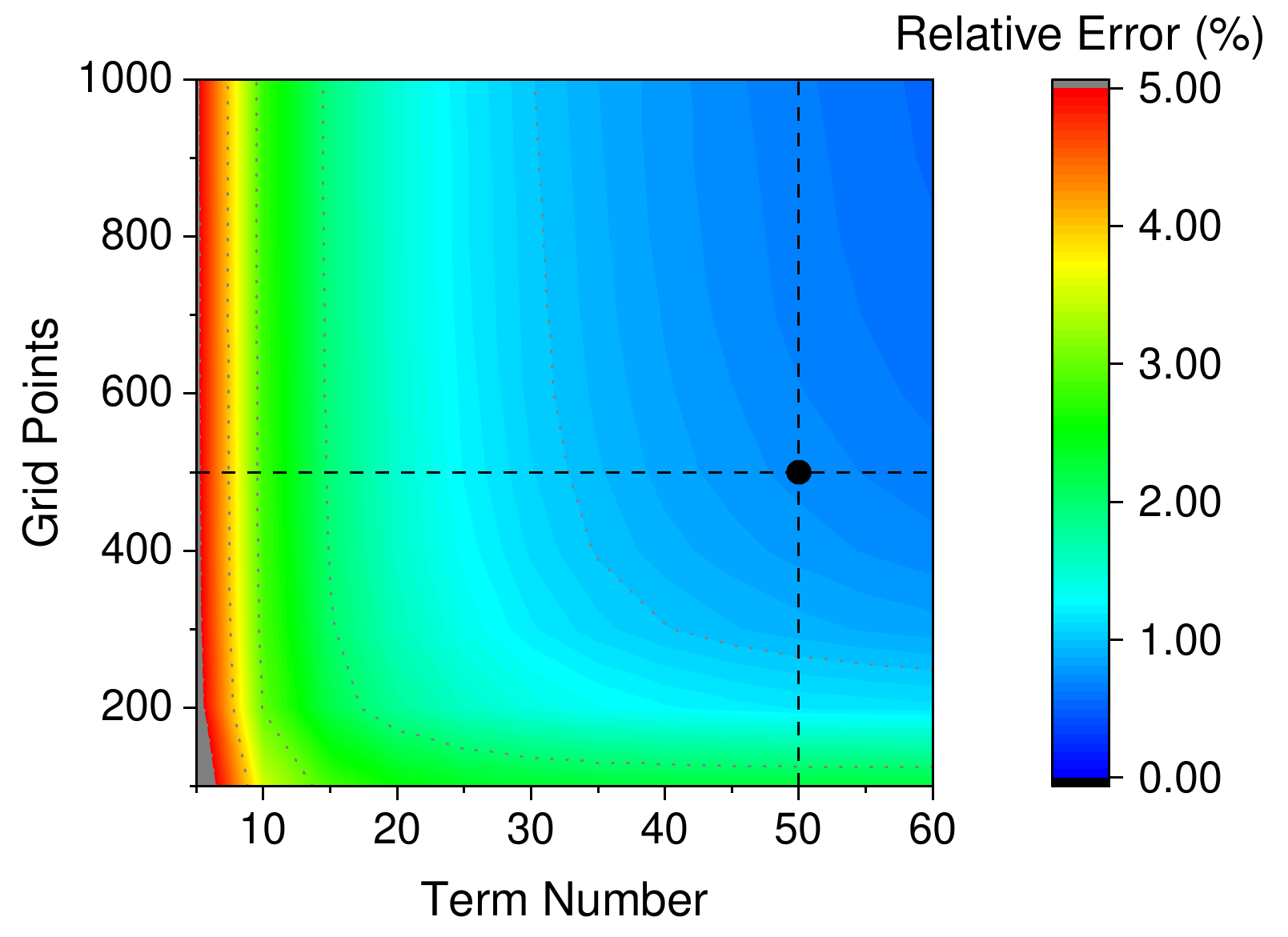}
    \caption{Relative errors of entropy by the SCF method with various parameters. Gray dotted lines signify integer percents, and the black point signifies parameters utilized in our reconstruction.}
    \label{fig:ConvTest}
\end{figure}

For the same situation, we calculate entropy using the SCF method with various parameters and then compare these results with the analytical results. The relative error obtained from the comparison is visualized in Fig.~\ref{fig:ConvTest}.

Based on the analysis of entropy, our approach demonstrates a high level of reliability. Opting for a denser grid and a larger basis set generally yields more accurate results, but at the expense of heightened computational time. Specificially, the time complexity of the term number and the grid size are $O(N^3)$ and $O(N)$. Consequently, striking the right balance between time efficiency and accuracy is of utmost importance. Given our available computational resources, all calculations presented in this paper were conducted employing 50 terms and 500 points.

\begin{figure}[!htbp]
    \centering
    \includegraphics[width=1\linewidth]{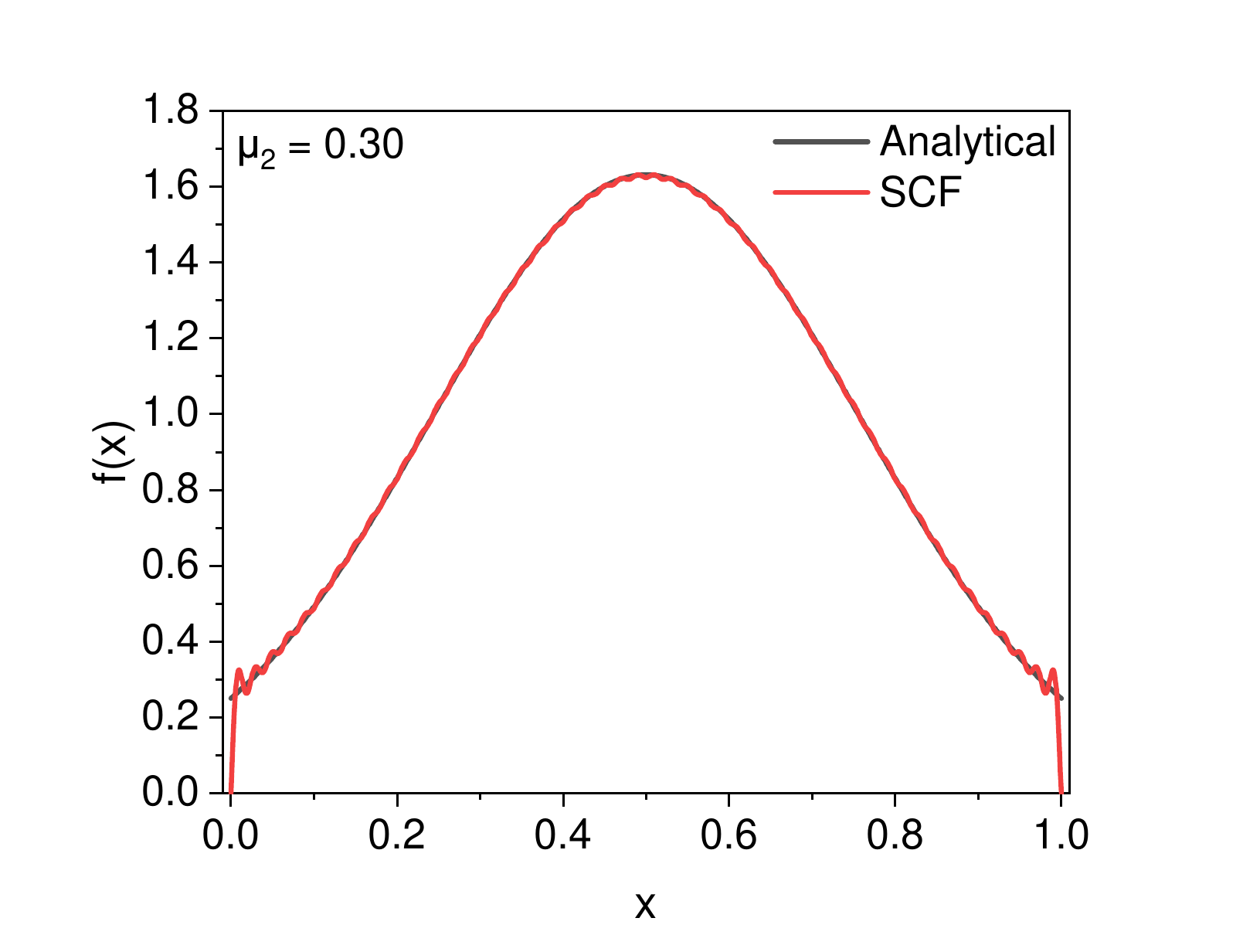}
    \caption{Comparison of the analytical and the SCF result.}
    \label{fig:N-result}
\end{figure}

The analysis discussed above is exclusively based on entropy, which only carries partial information about distribution. Therefore, a more nuanced comparison is now necessary. To elucidate the deviation at each data point, we compare the SCF result using the aforementioned parameters with the analytical result, which is shown in Fig.~\ref{fig:N-result}.

It becomes evident that the curve from the SCF method closely aligns with the Gaussian distribution derived through the variational approach. Notably, the SCF method produces the anticipated outcome within the majority of the function's range, indicating that our method is reliable enough. Nevertheless, slight oscillations can be observed at the tails of the distribution function. These oscillations primarily arise due to the nature of the basis set.

\subsection{Reconstruction Efficiency Assessment}

In this section, we engage in a discussion concerning a model that holds physical significance. Our objective is to assess the effectiveness of reconstruction under different numbers of constraints. The symmetric function presented below is a widely used form for characterizing the distribution of partons:
\begin{equation}
    f(x)=N_{\rho}\log(1+x^2(1-x)^2/\rho^2),
\end{equation}
where $N_{\rho}$ is the normalization constant.

This function is effective enough to describe PDF by adjusting $\rho$. Notably, there exists a unique correspondence between the parameters $\rho$ and $\mu_2$ within this distribution; for example, when $\mu_2$ equals $0.3$, the corresponding value of $\rho$ is $0.0658$. Considering the specific physical meanings of $\mu_2$, in this section, we apply $\mu_2$ to represent different functions.

\begin{figure}[!htbp]
    \centering
    \includegraphics[width=1\linewidth]{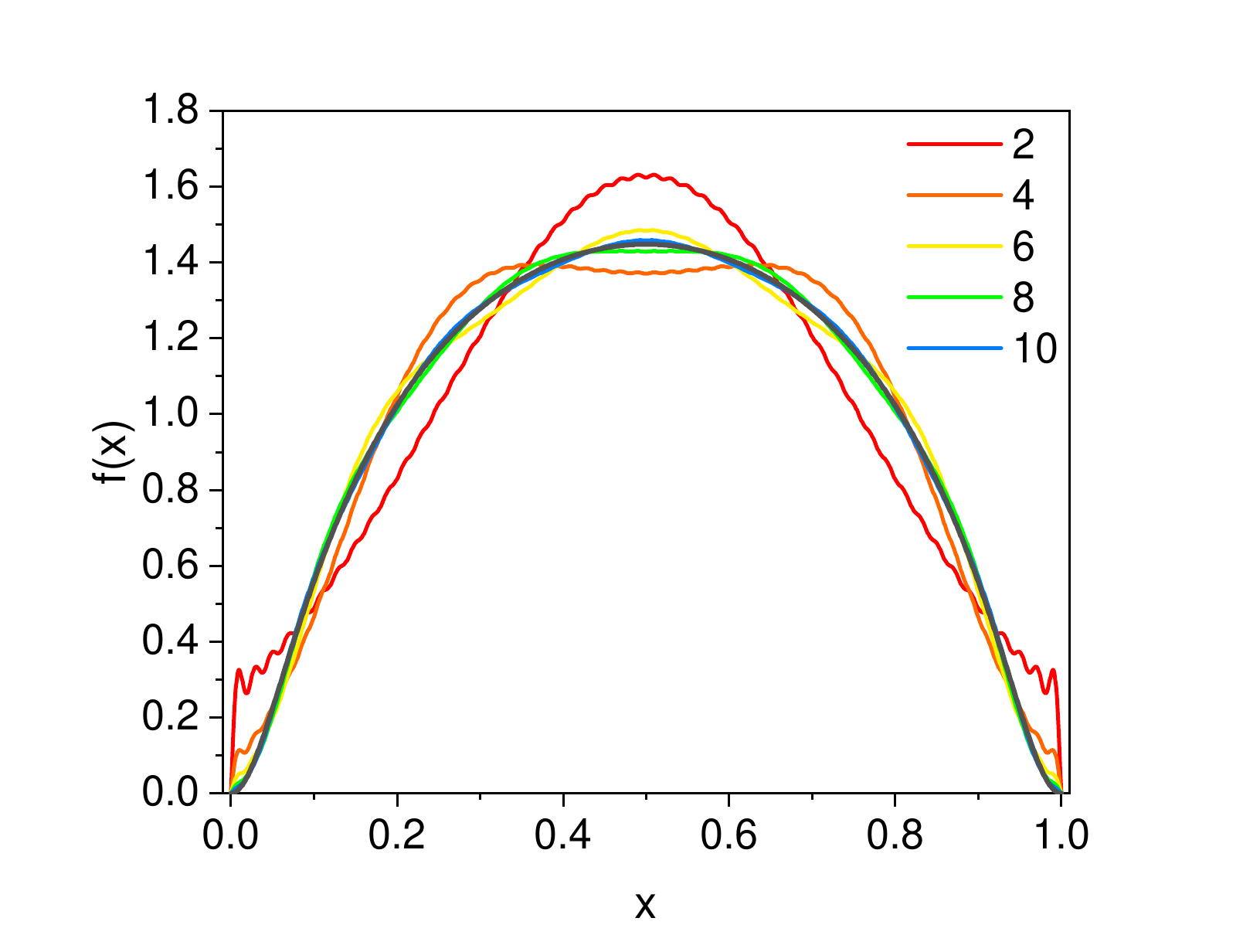}
    \caption{Variation in reconstruction results under different numbers of constraints. The black curve is the target symmetric function.}
    \label{fig:different-order-result-3}
\end{figure}

\begin{figure}[!htbp]
    \centering
    \includegraphics[width=1\linewidth]{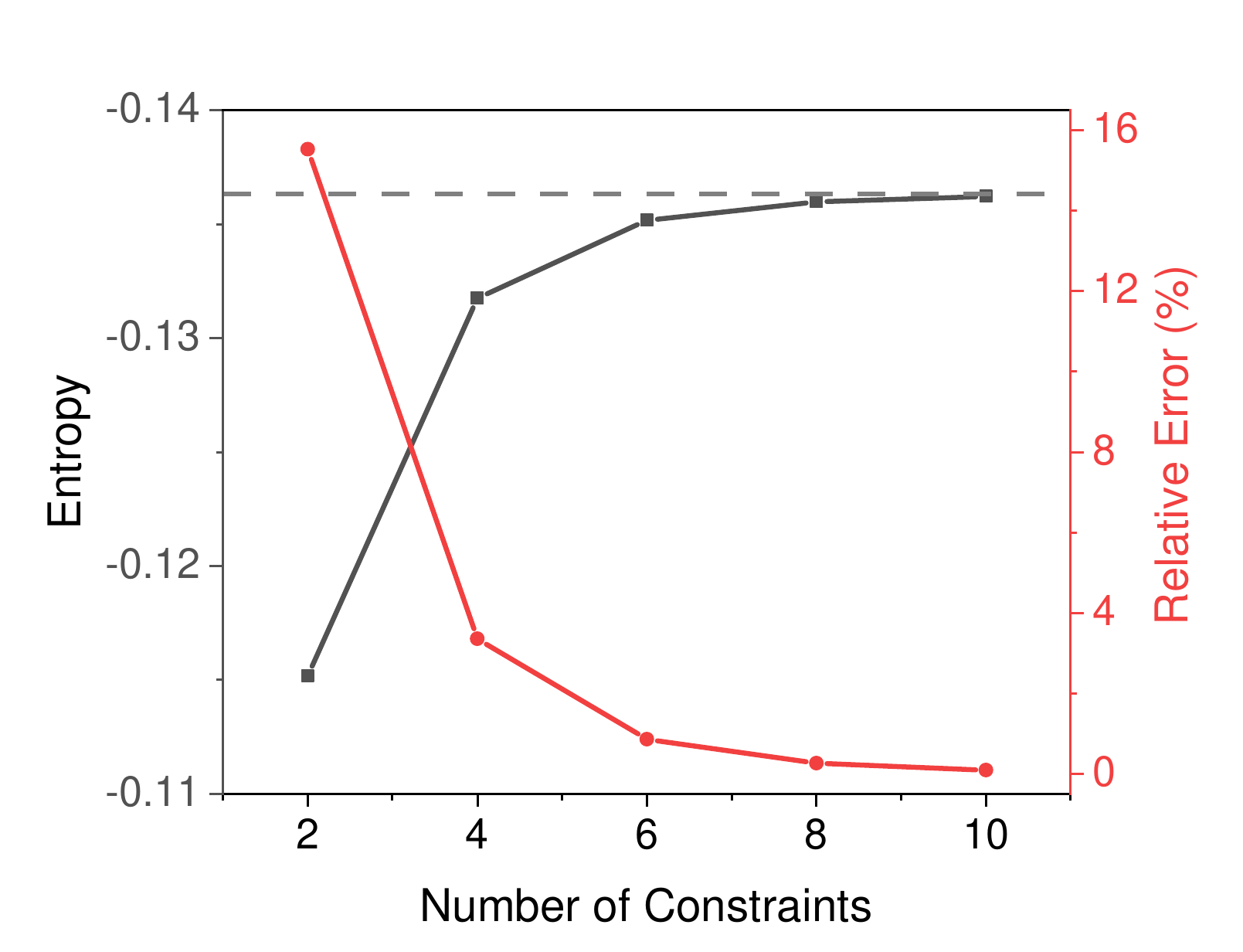}
    \caption{Variation in entropy under different numbers of constraints. Entropy is represented by the black curve, and relative error by the red curve. The gray dashed line is the entropy of the target symmetric function.}
    \label{fig:different-order-result-4}
\end{figure}

Subsequently, in order to show the efficiency of reconstruction under different numbers of constraints, we use the first few moments corresponding to $\mu_2=0.3$ as constraints to attempt to reconstruct the distribution. The resulting distribution functions and entropy are displayed in Figs. ~\ref{fig:different-order-result-3} and \ref{fig:different-order-result-4}, respectively. The reason only even numbers are considered is that odd-order moments are not independent because of the symmetry.

Figs.~\ref{fig:different-order-result-3}, \ref{fig:different-order-result-4} unmistakably illustrate that as additional constraints are integrated, the outcomes progressively approach the sought-after symmetric function, i.e., the efficiency of reconstruction becomes better. Importantly, it becomes evident that with a minimum of six moment constraints, the precision of the reconstruction reaches a high level, with an error margin of approximately $1\%$. Notably, employing ten constraints yields a notably robust and accurate reconstruction. Therefore, reconstruction with at least six constraints is suggested for a reliable result.

\begin{figure}[!htbp]
    \centering
    \includegraphics[width=1\linewidth]{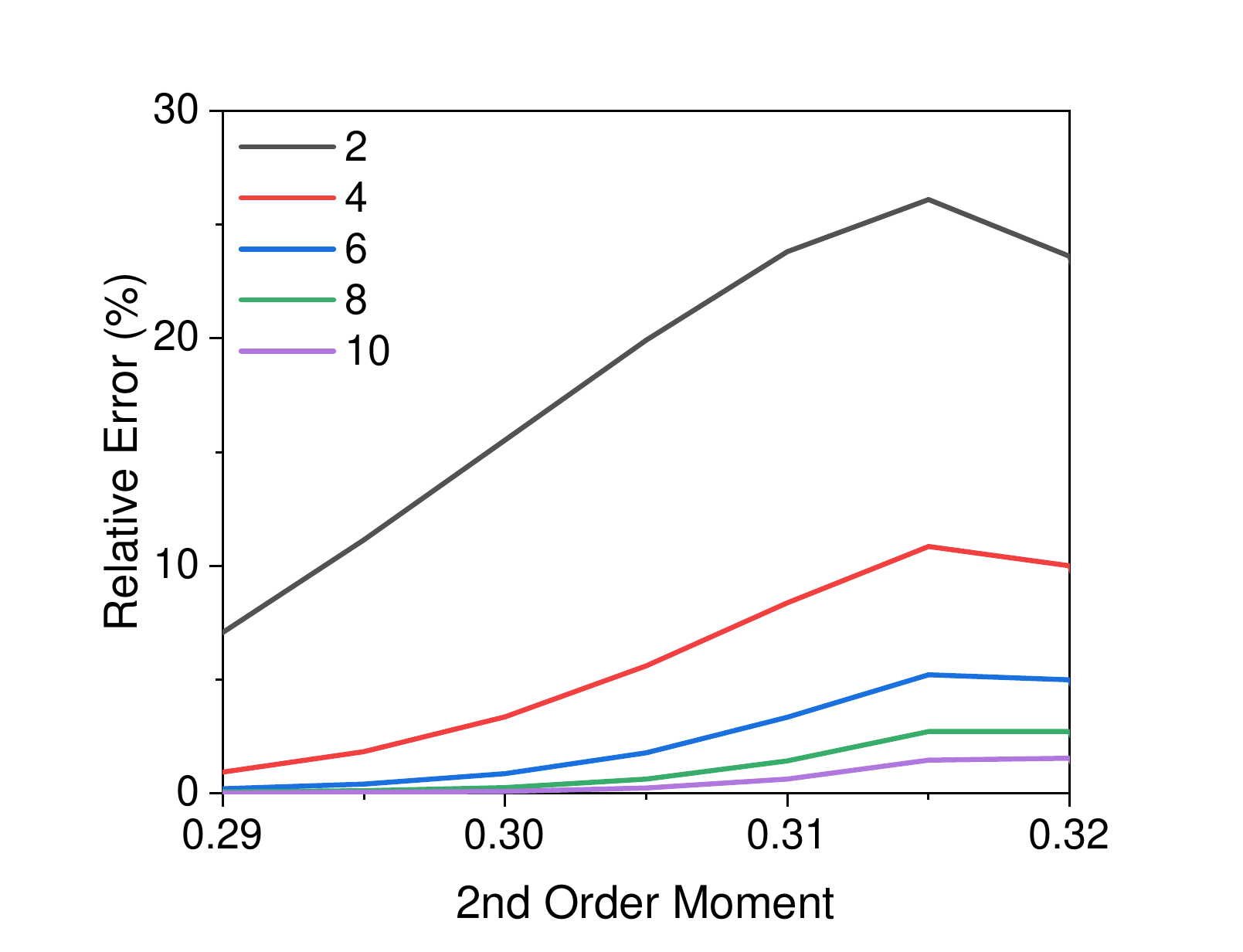}
    \caption{Variation in the relative error of entropy with the second-order moment.}
    \label{fig:different-order-result-5}
\end{figure}

It is crucial to emphasize that the comprehensive analysis thus far exclusively concerns the scenario wherein $\mu_2$ holds a value of $0.3$. Therefore, the calculations must be extended across a spectrum of $\mu_2$ values to validate the robustness of our previous conclusions. Fig.~\ref{fig:different-order-result-5} diligently presents the relative errors of entropy for varying $\mu_2$ values ranging from $0.29$ to $0.32$. This scope comprehensively covers the possible scenarios for PDF. This meticulous examination is undertaken to confirm the generalizability of the earlier-drawn conclusions.

The insights provided by Fig.~\ref{fig:different-order-result-5} indicate that relative errors exhibit an increasing trend as the second-order moment increases. However, for the six constraints case, the relative error remains confined within a threshold of $5\%$. This level of variability has been found to be satisfactory for a significant proportion of reconstruction tasks. Therefore, it can be reasonably concluded that a minimum of six constraints provides the necessary foundation for achieving reliable and accurate reconstructions.

\subsection{Reconstruction Based on Real Data}

This section is about the reconstruction based on real data with error bars. The first six moments of the pion valence-quark distribution are given using lattice QCD in Ref.~\cite{PhysRevD.104.054504}. However, these data are all at $5.2$ GeV, and the PDF at this scale is not symmetric. So we will evolve these data to the hadron scale according to the method in Refs.~\cite{Raya:2021zrz,wang2023sieving}, and the evolved results are shown in Table.~\ref{Moments from QCD}.
\begin{table}[!htbp]
    \centering
    \begin{tabular}{c|c c c c c c}
        \hline
        Order& 1& 2& 3& 4& 5& 6\\
        \hline
        Moment& $0.5$& $0.29(3)$& $0.19(5)$& $0.14(5)$& $0.10(5)$& $0.08(4)$ \\
        \hline
    \end{tabular}
    \caption{Data for the first six moments}
    \label{Moments from QCD}
\end{table}

\begin{figure}[!htbp]
    \centering
    \includegraphics[width=1\linewidth]{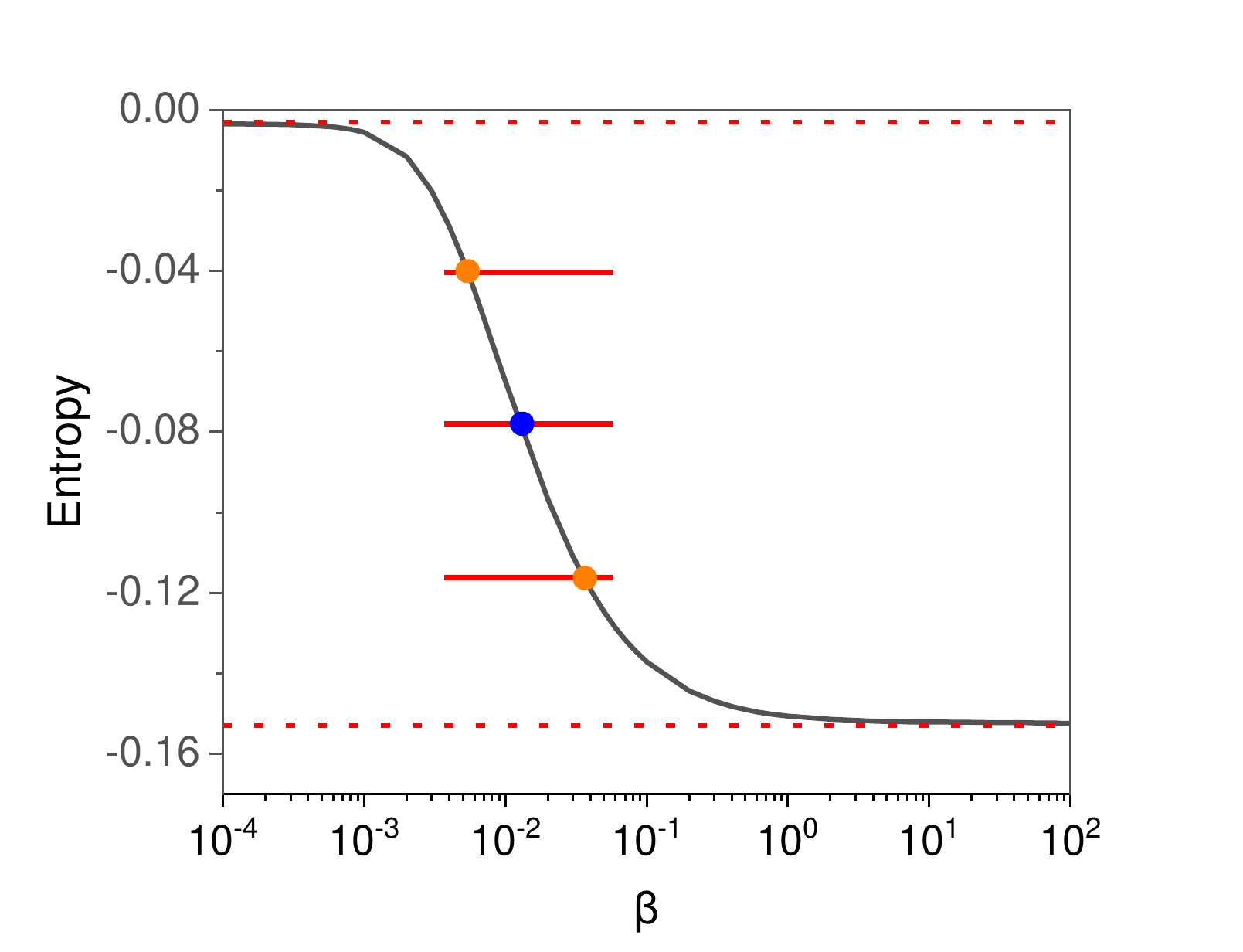}
    \caption{Variations in entropy values as $\beta$ changes. The axis is converted to a logarithmic scale. Red dashed lines serve as asymptotes, while the blue and orange dots correspond to the median and boundaries of the error range of $\beta$, respectively.}
    \label{fig:6}
\end{figure}

In order to select an appropriate value for $\beta$, it is crucial to understand how the entropy curve behaves under different $\beta$ settings. Fig.~\ref{fig:6} has been constructed to visually represent the changes in entropy with varying values of $\beta$.

\begin{figure}[!htbp]
    \centering
    \includegraphics[width=1\linewidth]{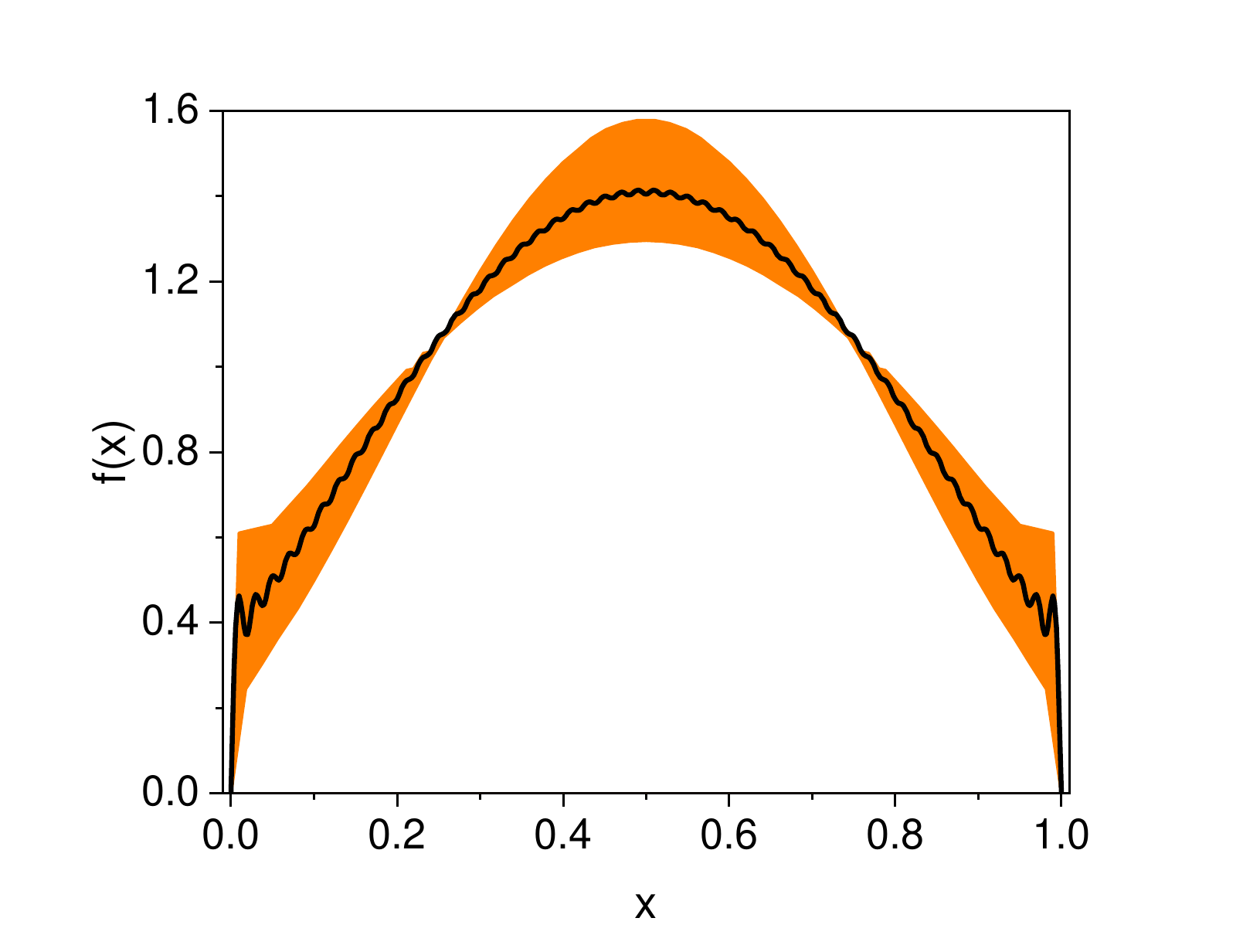}
    \caption{Reconstructed distribution function.}
    \label{fig:error bar-result-7}
\end{figure}

\begin{figure}[!htbp]
    \centering
    \includegraphics[width=1\linewidth]{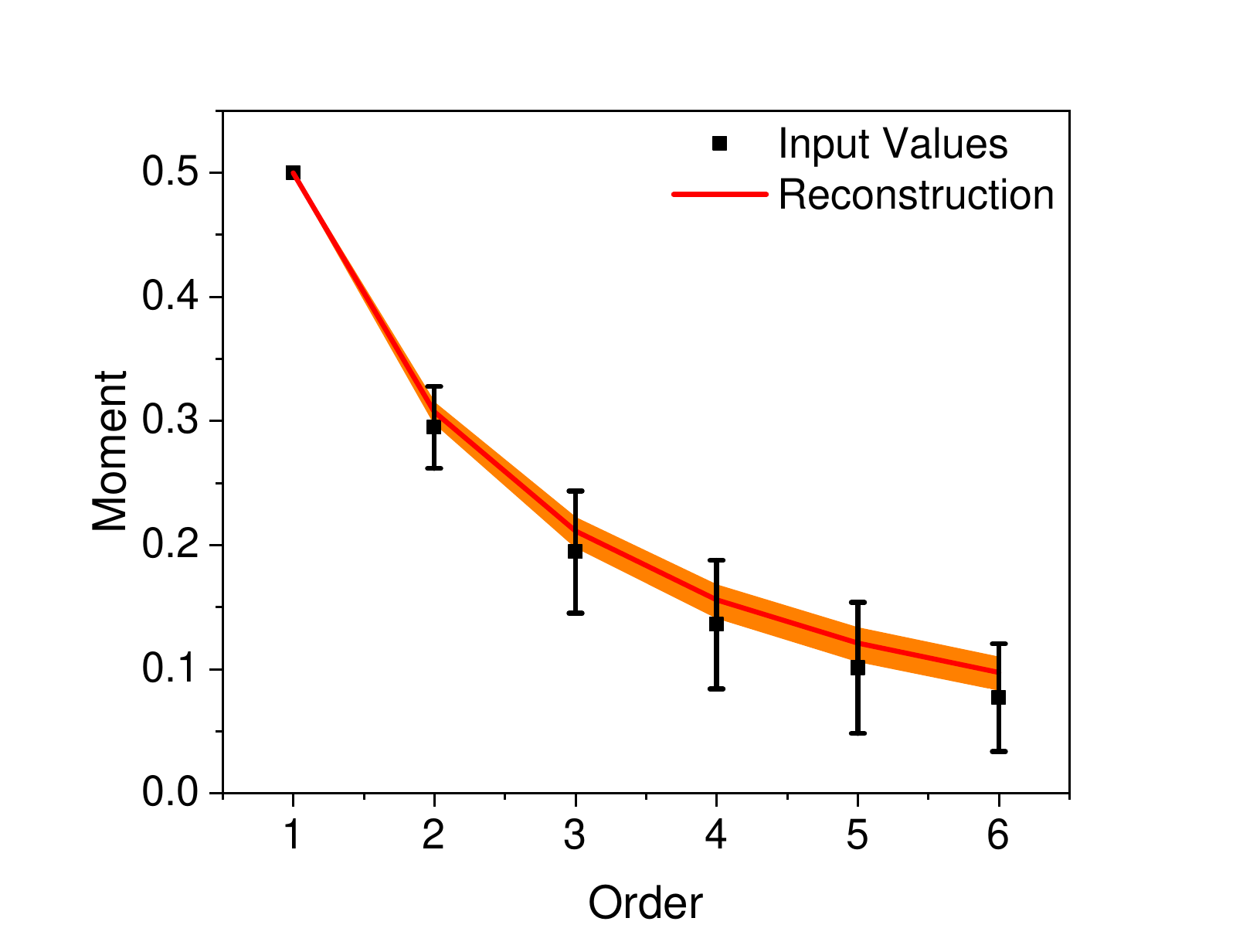}
    \caption{Comparison between the lattice input and moment error range of the reconstruction result.}
    \label{fig:error bar-result-8}
\end{figure}

In Fig.~\ref{fig:6}, we observe a significant trend: as $\beta$ increases, there is a consistent decrease in entropy. Notably, at both extremes, we observe a convergence towards two distinct values. One corresponds to a state where maximum entropy prevails, while the other represents a situation where constraints are rigorously enforced. This observation aligns seamlessly with our earlier discussions. As a consequence, it is reasonable to infer that the ideal value for $\beta$ lies somewhere between these two extremes. A straightforward approach is to select the midpoint between these extremes as the ideal value for $\beta$. To quantify our uncertainty, we recommend defining an error range by dividing the entropy range into four equal parts and designating the two central segments as the error range. This distribution function and its associated error range are described in Fig.~\ref{fig:error bar-result-7}. Additionally, in Fig.~\ref{fig:error bar-result-8}, we present a comparison between the lattice input and moment error range of the reconstruction result.

In Fig.~\ref{fig:error bar-result-7}, we present an outstanding reconstruction result, closely mirroring the findings reported in prior research as detailed in Ref.~\cite{PhysRevD.105.L091502}. This striking similarity underscores the robustness and consistency of our reconstruction method. Turning our attention to Fig.~\ref{fig:error bar-result-8}, it becomes evident that the error range of our reconstruction is well within compatibility bounds with the lattice input. However, it's important to note that our reconstruction's error range consistently lies slightly above the lattice input, particularly for higher-order moments, where a more noticeable deviation is observed.

In conclusion, the effectiveness and reliability of the proposed reconstruction method have been proven in this work. However, the pursuit of a more precise method for selecting the optimal $\beta$ value warrants further dedicated investigation and research.

\section{Summary}\label{4}
In this work, we combine the first few moments and entropy as constraints to define the Lagrange function and numerically reconstruct the PDF at the maximum point of the Lagrange function. To include the error in calculating moments with QCD, we replace the original moment constraints with Gaussian-shaped functions to soften the constraints. In this way, it is more natural and convincing than the methods that presuppose the function form of PDF artificially.

We comprehensively evaluate the convergence and reconstruction efficiency of this new method of reconstructing PDF. The evaluation results show that our method is reasonable. As the number of moments entered increases, the results become more accurate, and you can get high-quality reconstruction using only the first six moments as input. What's more, we select a set of lattice QCD results regarding moments in Ref.~\cite{PhysRevD.104.054504} as input to reconstruct the PDF. Finally, we can get an excellent reconstruction result and provide a reasonable error band.

With this PDF reconstruction method, which avoids artificial selection, reasonable and reliable results are obtained. And this method still has the potential for further development. For the input with error, we can multiply each $\Delta_i$ by $\beta_i$, which can increase the accuracy of the calculation but also increase the difficulty of the calculation. For the asymmetric case, we can replace the base set to handle it, but this will bring more computational complexity and potential convergence difficulty. It is hoped that this work can promote the research of PDF reconstruction in the field of hadron structure.

\begin{acknowledgments}
Work supported by Key Project for Undergraduate Teaching Reform and Quality Enhancement Research Plan in Ordinary Colleges and Universities in Tianjin(grant no. A231005505).
\end{acknowledgments}

\appendix

\nocite{*}

\bibliography{ref}% Produces the bibliography via BibTeX.

\end{document}